\documentclass{raa}

\usepackage{graphicx,times}             
\input{epsf.sty}                        

\newcommand{\asca}{{\it ASCA} }

\newcommand{\xmm}{{\it XMM-Newton} }

\newcommand{\chandra}{{\it Chandra} }

\newcommand{\ergs}{${\rm erg \ cm^{-2} \ s^{-1}}$ }

\newcommand{\erg}{${\rm erg \ s^{-1}}$ }

\begin{document}

   \title{The X-ray Nature of Nucleus in Seyfert 2 Galaxy NGC~7590
}

   \volnopage{Vol.0 (200x) No.0, 000--000}      
   \setcounter{page}{1}          

   \author{Xinwen Shu
      \inst{1}
   \and Junxian Wang
      \inst{1}
   \and Teng Liu
      \inst{1}
   \and Wei Zheng
      \inst{2}
   }

   \institute{
CAS Key Laboratory for Research in Galaxies and Cosmology, Department of
 Astronomy, University of Science and Technology of China, Hefei, Anhui 230026, China, xwshu@mail.ustc.edu.cn\\
        \and
Department of Physics and Astronomy, The Johns Hopkins University, { Baltimore, MD 21218, USA}\\   
}

   \date{Received~~2009 month day; accepted~~2009~~month day}

\abstract{ 
We present the result of the \chandra high-resolution observation of the 
Seyfert~2 galaxy NGC 7590.
This object was reported to show no X-ray absorption in the low-spatial resolution  
\asca data. 
The \xmm observations show that the X-ray emission of NGC 7590 is dominated 
by an off-nuclear {ultra-luminous X-ray source (ULX) and an} extended emission from the host galaxy, and the nucleus is 
rather weak, likely hosting a Compton-thick AGN. 
Our recent \chandra observation of NGC 7590 enables to remove the 
X-ray contamination from the ULX and the extended component effectively. 
The nuclear source remains undetected at $\sim4\times10^{-15}$ \ergs flux level.   
Although not detected, \chandra data gives a 2--10 keV flux upper limit 
of $\sim6.1\times10^{-15}$ \ergs (at $3\sigma$ level), a factor of 3 less than the 
\xmm value, strongly supporting the Compton-thick nature of the nucleus. 
In addition, we detected { five} off-nuclear X-ray point sources within the galaxy $D_{25}$ 
ellipse, 
all with 2 -- 10 keV luminosity 
above 2$\times$10$^{38}$ \erg (assuming the distance of NGC 7590). Particularly, 
the {ULX previously identified} by ROSAT data is resolved by \chandra into two distinct X-ray 
sources. Our analysis highlights the importance of high spatial resolution images in 
discovering and studying ULXs.
\keywords{
galaxies:active---galaxies:individual (NGC 7590)---galaxies:{nuclei}---X-rays:{galaxies}
}
}
   \authorrunning{Shu, Wang \& Liu et al. }            
   \titlerunning{X-ray Nature of Seyfert 2 Galaxy NGC 7590}  

   \maketitle

%
%
\section{Introduction}           
\label{sect:intro}

According to the current unification model for active galactic nuclei ({AGNs}), 
Seyfert 1 (Sy1) and Seyfert 2 (Sy2) galaxies are intrinsically 
same type of objects, and their 
observational differences are caused by orientation effects (Antonucci 1993).
In an Sy2 nucleus, the broad line region (BLR) is blocked by an optically thick torus 
along the line of sight, so that any broad emission lines (BELs) are not directly visible. 
The discovery of hidden BELs in many Sy2s from both near-IR spectroscopic and optical 
spectropolarimetry observations has given much support to this picture (e.g., 
Veilluex et al. 1997; 
Moran et al. 2000; Tran 2001; Shu et al. 2007, 2008). 
Further support for the unification model comes from the X-ray observations, 
showing that the column densities in Sy2s are typically above $10^{23}$ cm$^{-2}$ 
(see e.g., Risaliti, Maiolino \& Salvati 1999), much higher than those of Sy1s.  

However, recent observations have also questioned the applicability 
of the unification model to all AGN populations, finding that there exists a subset of ``{unobscured}" Sy2s that 
show no or very low X-ray absorption ($N_{\rm H}<10^{22}$ cm$^{-2}$, e.g. Pappa et al. 2001; 
Panessa \& Bassani 2002; Barcons et al. 2003; Walter et al. 2005; Gliozzi, Sambruna \& Foschini 2007).
The peculiar X-ray spectra of these ``{unobscured}" Sy2s could be explained by the absence of a BLR, 
where their appearance as Sy2s is intrinsic and not a result of {the X-ray} absorption (Nicastro et al. 2003; 
Georgantopoulos \& Zezas 2003; Bianchi et al. 2008; Panessa et al. 2009, Tran et al. 2011).
{ Alternatively, the appearance of the X-ray ``unobscured'' Sy2s could be due to} 
an extremely high dust-to-gas ratio compared with the Galactic value 
(see e.g., Huang et al. 2011). 

However, one has to be cautious in identifying the candidates of 
the ``unobscured" Sy2s, since their type 2 classification may be uncertain due to the 
insufficient optical spectroscopy quality (e.g., Panessa et al. 2009; Gliozzi et al. 2010). 
On the other hand, some of Sy2s that appear to lack {the X-ray} absorption may be indeed Compton-thick. 
In such sources where the intrinsic absorption is so high ($N_H>10^{24}$ cm$^{-2}$) that 
the direct component below 10 keV is completely absorbed, the $unabsorbed$ scattered component or 
the extended emission from the host galaxy would dominate the observed spectrum in the 2-10 keV band 
(see e.g., Pappa et al. 2001). 
Brightman \& Nandra (2008) presented a detailed spectral
analysis of six {unabsorbed Seyfert 2 candidates}, and found {that} four
out of them are in fact heavily obscured.
Furthermore, Shi et al. (2010) presented a multi-wavelength study of a sample of ``unobscured" Seyfert 2 galaxies,
and found that most of them are actually intermediate-type AGNs with weak {BELs} or 
Compton-thick sources. 

Recently, with three \xmm observations we carried out a preliminary X-ray study of NGC 7590, a Sy2 previously identified to be unabsorbed in the X-ray (Shu, Liu, \& Wang 2010, hereafter Paper I).
 We found that the X-ray 
emission of NGC 7590 is dominated by an off-nuclear {ultra-luminous X-ray source (ULX) and an} extended emission from the host galaxy. 
The small ratio of the 2--10 keV to the {[O {\sc iii}]} fluxes suggests that 
this galaxy is likely Compton-thick rather than X-ray ``unobscured" as previously thought. 
However, due to the contamination from the ULX and the extended component, we are unable to
isolate the nuclear X-ray emission for NGC 7590. 
In this paper, we investigate the X-ray nature of nucleus in NGC 7590, using higher spatial resolution 
\chandra observation. The new \chandra data has enabled a clear view of the true 
nuclear emission. Although not detected, 
the derived flux upper limit {in the 2--10 keV} is a factor of 3 less than 
{that measured by the \xmm}, confirming the Compton-thick nature 
of the {NGC 7590} nucleus.  
  

\section{Observations}
\label{sect:Obs}

NGC 7590 was observed by \chandra on 2010 August 22 (observation ID 12240, PI: Shu) for an exposure 30 ks, 
using the front-illuminated chips of the Advanced CCD Imaging Spectrometer (ACIS-I). 
The galaxy was placed at the aimpoint of the I3 chip, which provides 
a spatial resolution of 0.492\arcsec. 
The filed of view in this mode covers the whole galaxy (encompasses the full $D_{25}$
\footnote{$D_{25}$ is the apparent major isophotal diameter measured at the surface 
brightness level $\mu_B$=25.0 mag/arcsec$^2$.} area of the galaxy). 
The data were processed with the CIAO (version 4.3) and CALDB (version 4.4.1), following 
standard criteria. 
Level 2 event lists were reprocessed with observation-specific bad pixel files. 
The CIAO task \textit{wavedetect} was run to determine the detections of the {X-ray source candidates} and 
the {resulting} source positions were used for the following spectra extraction. 




%

\section{Data analysis}
\label{sect:analysis}
 \subsection{Imaging}

   \begin{figure}
   \centering
   \includegraphics[width=10cm, angle=0]{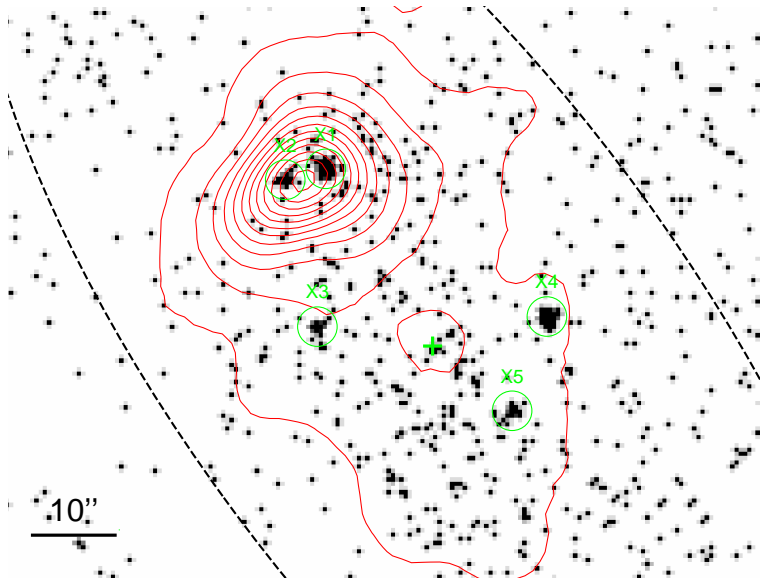}
   \caption{\chandra image of NGC 7590 with {the} \xmm X-ray contours (red) overlaid.
The cross marks the optical nucleus of the galaxy. The green circles correspond to
five X-ray sources detected by \chandra inside {the galaxy} $D_{25}$ ellipse (the black dotted line).
Thanks to its sub-arcsecond spatial resolution, \chandra clearly reveals two closely-spaced X-ray sources
(X1 and X2), which were not resolved with the \xmm images. }
   \label{Fig:demo1}
   \end{figure}

There are totally five sources detected by \chandra inside the {galaxy $D_{25}$ ellipse}. The detected sources, 
together with the \xmm X-ray contours, are shown in Figure 1. The dashed line represents the $D_{25}$ 
ellipse of NGC 7590,  
{while the cross marks the position of the galaxy optical nucleus. 
As can be seen from Figure 1, the nuclear source is not detected by \chandra,} 
for which the $3\sigma$ upper limit of counts is 20.6, 
calculated following Gehrels (1986) for Poisson statistics.
It is interesting to note that the brightest point source, which was detected by \xmm about 25\arcsec away 
from the galaxy nucleus (also originally identified in the $ROSAT$ All-Sky Survey as an 
ULX, see Colbert 
\& Ptak 2002), was resolved into two sources X1 and X2 with the high resolution \chandra image.  

The details of all the sources above are given in Table 1. In addition to the source name and 
the equatorial coordinates (J2000.0), we give the counts, absorbed flux, and luminosity (all in the 2-10 
keV band) in columns (3), (4) and (5), respectively. For X1, X2 and X4, source counts were 
extracted from a circular aperture centered on the \textit{wavdetect} source position, with a radius 
of 4.67 pixels (or 2.3\arcsec, 1.3 times the on-axis 95\% encircled energy radius at 1.5 keV on ACIS-I). 
Background counts were taken from an annulus with an inner radius of twice (or 3 times for X1 and X2), 
and an outer radius of 3.5 times (or 5.5 times for X1 and X2) the source circle radius, avoiding the nearby point sources that could fall within the annulus. 
For the remaining two off-nuclear sources, X3 and X5, which have no enough counts 
(less than 20 counts) for a meaningful  
spectra fitting, we 
performed a conversion from count rate to flux assuming a power-law spectrum of $\Gamma=1.9$, 
consistent with the spectrum of low-mass X-ray binaries in nearby galaxies 
(see, e.g., Prestwich et al. 2003). 
\begin{table}
\begin{center}
\caption[]{NGC 7590 nucleus and the detected X-ray sources.}\label{Tab:publ-works}

\begin{tabular}{lcccc}
\hline\noalign{\smallskip}
\hline\noalign{\smallskip}
 Name  &  Equatorial Coordinates & Net Counts  & $F_{\rm 2-10~keV}$ & $L_{\rm 2-10~keV}$ \\
       &    (J2000)             &  & $10^{-14}$ \ergs & $10^{39}$ \erg \\
\hline\noalign{\smallskip}
 Nucleus & 23 18 54.8,   -42 14 21 & $<20.6$  & $<0.61$  & $<0.37$ \\
 X1      & 23 18 55.9,   -42 14 00 & $77.5\pm8.9$ & 2.8$^{+1.3}_{-0.8}$ & 1.7$^{+0.8}_{-0.5}$\\
 X2      & 23 18 56.5,   -42 14 01 & $32.4\pm5.8$ & 0.86$^{+0.78}_{-0.20}$& 0.53$^{+0.41}_{-0.13}$ \\
 X3      & 23 18 56.0,   -42 14 18 & $17.2\pm4.4$ & 0.47$^\dag$& 0.28$^\dag$ \\
 X4      & 23 18 53.6,   -42 14 17 & $61.0\pm7.9$ & 1.11$^{+0.63}_{-0.47}$ & 0.68$^{+0.35}_{-0.31}$ \\
 X5      & 23 18 54.0,   -42 14 28 & $16.3\pm4.5$ &0.44$^\dag$ & 0.27$^\dag$\\
\noalign{\smallskip}\hline
\end{tabular}
\end{center}
$^\dag$: The flux and luminosity are estimated by assuming a power-law spectrum {with}
 $\Gamma=1.9$ (see Section 3.1 for details).
\end{table}

\subsection{Spectroscopy}

{ Using the extraction radius mentioned above, we obtained spectra 
for three brightest sources (X1, X2 and X4) for which it is possible to 
perform spectral analysis.}  
The spectra were then grouped to have at least 1 count per bin, and the $C$-statistics (Cash 1979) 
was adopted for minimization. Spectral fitting was performed in the 0.3-7 keV range using XSPEC 
(version 11.3.2). All statistical errors given hereafter correspond to  
90\% confidence for one interesting parameter ($\Delta \chi^2=2.706$), 
unless stated otherwise. In all of the model fitting, the 
Galactic column density was fixed at $N_{\rm H} = 1.96 \times
10^{20} \rm \ cm^{-2}$ (Dickey \& Lockman 1990). 
All model parameters will be referred to in the source frame.  
We present in Table 2 the results of spectral fits for the above three sources.  

\begin{table}
\begin{center}
\caption[]{ Results of X-ray spectral fitting.}\label{Tab:publ-works}
 \begin{tabular}{clcl}
  \hline\noalign{\smallskip}
  \hline\noalign{\smallskip}
 Target  &  $N_{\rm H}$ (10$^{22}$ cm$^{-2}$)   &  $\Gamma$  &  $C/dof$ \\
 \hline\noalign{\smallskip}
X1  & $0.5^{+0.5}_{-0.4}$ & 1.96$^{+0.85}_{-0.69}$ & 46.1/63 \\
X2  & $<0.8$              & 1.91$^{+1.03}_{-0.78}$ & 25.3/27 \\
X4  & $<0.4$              & 2.23$^{+0.74}_{-0.60}$ & 31.8/53 \\
  \noalign{\smallskip}\hline
\end{tabular}
\end{center}
\end{table}

All three sources are reasonably well fitted with a simple absorbed power-law model. 
Although with substantial uncertainties due to the limited photon statistics, 
the photon index ($\Gamma \simeq2.0^{+0.8}_{-0.7}$) for the brightest source X1 is consistent with 
the value measured by \xmm data. 
{ Note that due to the lower spatial resolution of \xmm, the resulting spectra  
are in fact consisting of the summed emission from both X1 and X2. } 
The spectra and residuals of X1 and X2 are shown in Figure 2. 

The third source, X4, 
has luminosity from \chandra spectral fits (assuming the distance of NGC 7590, see Table 1) 
exceeding the Eddington luminosity 
for stellar mass X-ray binaries of $2\times10^{38}$ \erg (Makishima et al. 2000). 
This object was undetectable in previous $ROSAT$ HRI observations (Liu \& Bregman 2005), and 
 recent \xmm observations show only weak detections due to the contamination from the host galaxy.  
If this object could be associated with {an} ULX, it is interesting to {investigate} whether the \chandra detection 
corresponds to a recurrence or {an} outburst of ULX (see e.g., Bauer et al. 2005). 
We present the light curve of X4 in Figure 3, taken from $ROSAT$ (circles), 
\xmm (triangles), and \chandra (square). 
Although there appears a long-term brightness variability, the upper limits of the flux prevent us 
drawing any conclusive claims.

%
   \begin{figure}
   \centering
   \includegraphics[width=16cm, angle=0]{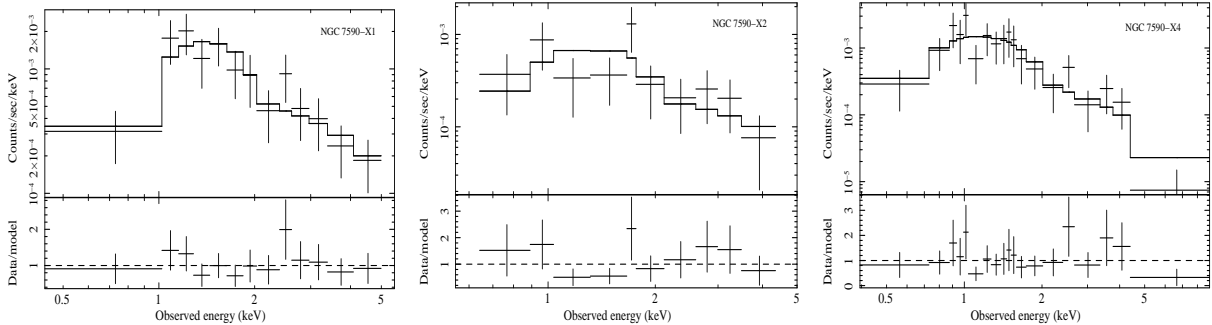}
   \caption{
\chandra spectra of {three} off-nuclear X-ray sources in NGC 7590, together with the best-fit 
{model and residuals}. 
}   \label{Fig:demo1}
   \end{figure}

%
   \begin{figure}
   \centering
   \includegraphics[width=12cm, angle=0]{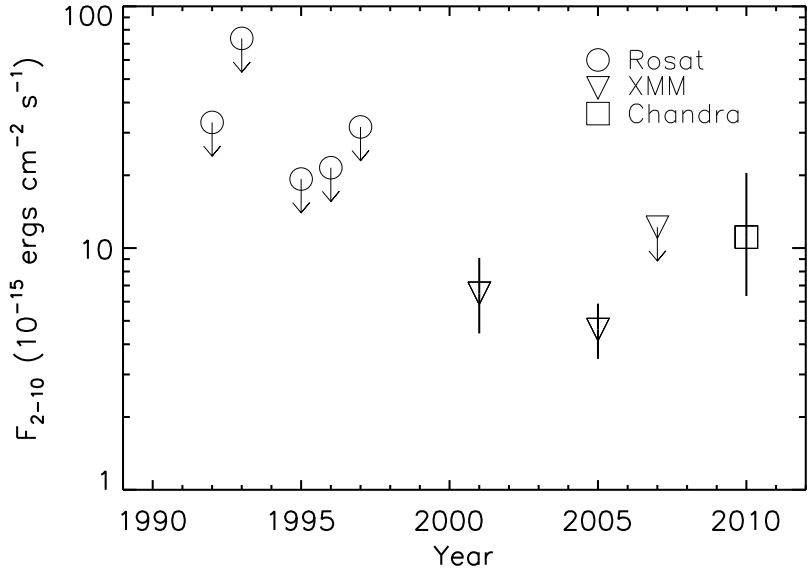}
   \caption{Light curve of NGC 7590--X4. Data are taken from $ROSAT$ (circles), \xmm (triangles), and \chandra 
(squares). The statistical errors on the flux shown correspond to 90\% confidence. Arrows represent the 
corresponding $3\sigma$ upper limit for fluxes. } 
   \label{Fig:demo2}
   \end{figure}
\begin{figure}
   \vspace{0.5cm}
   \begin{center}
   \plotone{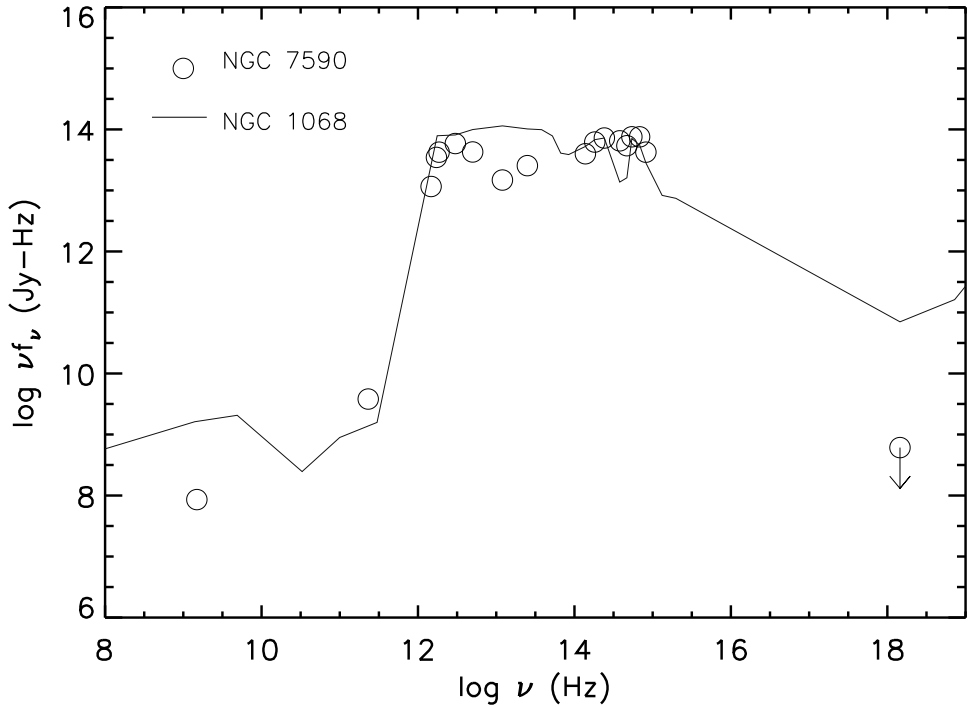}
   \caption{
   The NGC 7590 SED (open circles). The upper limit of the X-ray flux is given by \chandra observation, while 
the fluxes at other bands are taken from NED. For comparison we also show the SED of the 
archetypal Compton-thick Seyfert 2 galaxy NGC 1068 (solid line).   
The NGC 7590 SED was normalized to match that of NGC 1068 at optical $i$-band.
}\label{Fig:plot1}
   \end{center}
\end{figure}

\section{Discussion}
\label{sect:discussion}

NGC~7590 was previously identified to be an ``unobscured" Seyfert 2 galaxy, 
based on the poor X-ray spatial resolution \asca observation ($\sim$1$\arcmin$) 
and spectrum (Bassani et al. 1999). 
The higher spatial resolution ($\sim$6$\arcsec$ PSF FWHM) \xmm observations show that 
the X-ray flux of NGC 7590 nucleus is contaminated by a nearby bright ULX and {an} extended 
component from the host galaxy (see Paper I). 
 Because of the strong contamination, 
{ the \xmm data can only give an upper limit of the nuclear X-ray 
emission with $F_{2-10~\rm keV}$ = 1.6$\times10^{-14}$ \ergs.
 From the derived $T$ ratio ($F_{2-10~\rm keV}/F_{\rm [O~III]}<$0.09, 
a value in the range for Compton-thick AGNs, see e.g., Guainazzi et al. 2005), we 
conclude that NGC 7590 likely hosts a heavily obscured nucleus (Paper I).} 
Our new \chandra observation enables to remove the X-ray contaminations effectively 
and provides the direct X-ray view to the NGC 7590 nucleus. 
The \chandra data show that the NGC 7590 nucleus is rather weak, with a 2-10 keV flux 
upper limit of 0.6$\times10^{-14}$ \ergs. 
{Although not detected, the corresponding upper limit on the $T$ ratio ($<$0.033) 
suggests that the obscuration towards the nucleus is likely 
Compton-thick rather than ``unobscured" as previously thought, supporting the results
 of the \xmm observations. }


Another insight into the {X-ray} nature of the NGC 7590 nucleus can come from the energy production mechanisms at different 
wavebands of the galaxy. Figure 4 shows the SED of NGC 7590 (open circles), 
constructed using the data collected from the NASA NED. 
The SED for NGC 7590 
is compared with that of NGC 1068 (solid line), an archetypal Compton-thick Seyfert 2 galaxy (e.g., Pounds et al. 2006). 
Given the scaling of the comparison SEDs (normalized at optical $i$-band), 
the NGC 7590 SED multiwavelength data are in general 
agreement with an obscured AGN template. 
However, there is a disagreement of emission in the radio and near-IR band, which could be 
due to a more intense star-formation in NGC 1068. 
On the other hand, it is evident that NGC 7590 is relatively weak in the X-ray, 
which could be explained by a combination of both low accretion central black 
hole and strong absorption of the nuclear emission.
If NGC 7590 hosts a Compton-thick AGN with a SED similar to that of NGC 1068, then a 
luminous X-ray source should be present at higher energies ($\sim$20-30 keV), which 
could be detected by future hard X-ray imaging telescopes (i.e., NUSTAR and ASTRO-H). 
However, caution must be kept in mind in interpreting the SED, as the measurements have been 
taken at different epochs and {used} different apertures, in which the contamination from 
the host galaxy could be important.

With the unprecedented sub-arcsecond spatial resolution of \chandra, it is possible to separate even very closely 
spaced point sources, and easily distinguish them from surrounding diffuse emission. 
The \chandra image of NGC 7590 clearly shows two sources (labeled X1 and X2 in this paper), 
about 25 arcsec north-east from the position 
of the optical nucleus (see Figure 1). The two sources were not separated by 
lower resolution $ROSAT$ HRI and \xmm observations, and have previously been 
identified as an ULX with 2-10 keV luminosity 
$\sim5.7\times10^{39}$ \erg (Colbert \& Ptak 2002; Paper I). 
The \chandra spectra of both sources can be adequately fitted with a simple absorbed power law (see Table 2), 
though the parameters were loosely constrained due to the poor statistics. 
Note that the results of spectral analysis for both sources are consistent with each other, 
but with X1 showing 
a factor of {3} higher flux than X2. 
As shown in Table 1, by simply summing the flux from X1 and X2, we found a 2-10 keV luminosity of $\simeq3.7\times10^{39}$ \erg, a factor of $\sim2.5$ lower 
than what was reported by Paper I for the \xmm observations. 
Given the possible contamination by the emission from host galaxy and larger extraction 
radius of \xmm data, we cannot tell whether there is any variability of the ULX flux 
in the \chandra spectrum.  

Our analysis highlights the importance {of} utilizing the \chandra to discover and study ULXs. 
{The sub-arcsecond spatial resolution of \chandra is }essential to confirm the point-like nature of ULX candidates, and perhaps resolve additional sources that 
modest angular resolution observatories (i.e., \xmm) could not.  
Such confusion problems should be taken into account when studying the statistical properties of 
the X-ray source populations, in particular in distant galaxies with limited spatial 
resolution of X-ray observations, {where} the ULX luminosity may be \textit{overestimated}.
Since the X-ray luminosity is a defining property of the ULXs, the question we could ask instead is 
whether ULXs {represent} the high-luminosity end of a continuous distribution of typical X-ray 
sources such as X-ray binaries (e.g., Grimm et al. 2003; Swartz et al. 2004; Liu, Bregman \& Irwin 2006),
 or they may include new classes of objects including intermediate-mass 
black holes (Colbert et al. 2004; Farrell et al. 2009; Swartz et al. 2011). 
With the \chandra high resolution observations, we may need to re-visit the correlation between ULXs and 
star formation (e.g., Swartz et al. 2004; Liu et al. 2006; Walton et al. 2011), 
and to search for further similarities and/or differences between ULXs and less-luminous sources 
in both spiral and elliptical galaxies to confirm or rule out their X-ray binary nature.



\begin{acknowledgements}
X.W.S. thanks the support from China postdoctoral foundation.
This work has been supported by the Chinese National Science Foundation
(Grant No. 10825312, 11103017), the Fundamental Research Funds for the Central Universities
(Grant No. WK2030220004, WK2030220005).
{Partial support for this work was provided by NASA through Chandra Award GO1-12134X.} 
The authors are grateful to the \chandra
instrument and operations teams for making the observation
possible.
\end{acknowledgements}


\begin{thebibliography}{99}
\bibitem[Ant1993]{ant1993}
Antonucci, R. 1993, \araa, {31, 473}
\bibitem[Barcons2003]{Bar03}
Barcons X., Carrera F. J., Ceballos M. T., 2003, MNRAS, 339, 757

\bibitem[Bassani1999]{Bas99}
Bassani, L., Dadina, M., Maiolino, et al. 1999, ApJS, 121, 473

\bibitem[Bauer]{Bau05}
Bauer, M., \& Pietsch, W., 2005, A\&A, 442, 925

\bibitem[Bianchi2008]{Bia08}
Bianchi S., Corral A., \& Panessa F., et al. 2008, MNRAS, 385, 195
\bibitem[Brightman2008]{Bri08}
Brightman M., Nandra K., 2008, MNRAS, 390, 1241
\bibitem[Cash1979]{Cash79}
Cash, W. 1979, ApJ, 228, 939
\bibitem[Colbert2002]{Col02}
Colbert, E. J. M., \& Ptak, A. F. 2002, \apjs, 143, 25
\bibitem[Colbert2004]{Col04}
Colbert E. J. M., Heckman T. M., Ptak A. F., Strickland D. K., Weaver K. A., 2004, ApJ, 602, 231
\bibitem[Dickey \& Lockman 1990]{dick1990}
Dickey, J. M., \& Lockman, F. J. 1990, ARA\&A, 28, 215
\bibitem[Farrell2009]{Far09}
Farrell S. A., Webb N. A., Barret D., Godet O., Rodrigues J. M., 2009, Nature, 460, 73

\bibitem[Gehrels(1986)]{1986ApJ...303..336G} Gehrels, N.\ 1986, \apj, 303, 
336 

\bibitem[Georgantopoulos2003]{Geo03}
Georgantopoulos, I., \& Zezas, A. 2003, ApJ, 594, 704
\bibitem[Gliozzi2007]{Gli07}
Gliozzi M., Sambruna R. M., Foschini L., 2007, ApJ, 662, 878

\bibitem[Gliozzi et al.(2010)]{2010ApJ...725.2071G} Gliozzi, M., Panessa, 
F., La Franca, F., et al.\ 2010, \apj, 725, 2071 

\bibitem[Grimm2003]{Gri03}
Grimm H., Gilfanov M., Sunyaev R., 2003, MNRAS, 339, 793


\bibitem[Gua2005]{Gua05}
Guainazzi, M., Matt, G., \& Perola, G. C. 2005, A\&A, 444, 119 

\bibitem[Huang et al.(2011)]{2011ApJ...734L..16H} Huang, X.-X., Wang, 
J.-X., Tan, Y., Yang, H., \& Huang, Y.-F.\ 2011, \apjl, 734, L16 

\bibitem[liu2006]{Liu06}
Liu J., \& Bregman J. N., 2005, ApJS, 157, 59

\bibitem[liu2006]{Liu06}
Liu J., Bregman J. N., Irwin J., 2006, ApJ, 642, 171 

\bibitem[Moran2000]{Mor00}
Moran, E. C., Barth, A. J., Kay, L. E.,
\& Filippenko, A. V. 2000, ApJ, 540, L73

\bibitem[Mak]{Mak00}
Makishima, K., Kubota, A., Mizuno, T., et al. 2000, ApJ, 535, 632

\bibitem[Nicastro2003]{Nic03}
Nicastro, F., Martocchia, A., \& Matt, G. 2003, ApJ, 589, L13
\bibitem[Panessa2002]{Pan02}
Panessa, F., \& Bassani, L. 2002, A\&A, 394, 435
\bibitem[Panessa2009]{Pan09}
Panessa, F., Carrera, F. J., Bianchi, S., et al. 2009, \mnras, 398, 1951
\bibitem[Pappa2001]{Pap01}
Pappa, A., Georgantopoulos, I., Stewart, G. C., \& Zezas, A. L. 2001,
MNRAS, 326, 995
\bibitem[Pounds2006]{Pou06}
Pounds K., \& Vaughan S. 2006, MNRAS, 368, 707

\bibitem[Prestwich et al.(2003)]{2003ApJ...595..719P} Prestwich, A.~H., 
Irwin, J.~A., Kilgard, R.~E., et al.\ 2003, \apj, 595, 719 


\bibitem[Risaliti1999]{Ris99}
Risaliti G., Maiolino R., Salvati M., 1999, ApJ, 522, 157
\bibitem[Shi et al.(2010)]{2010ApJ...714..115S} Shi, Y., Rieke, G.~H.,
Smith, P., et al.\ 2010, \apj, 714, 115
\bibitem[Shu et al.(2007)]{2007ApJ...657..167S} Shu, X.~W., Wang, J.~X.,
Jiang, P., Fan, L.~L., \& Wang, T.~G.\ 2007, \apj, 657, 167

\bibitem[Shu et al.(2008)]{2008ChJAA...8..204S} Shu, X.-W., Wang, J.-X., 
\& Jiang, P.\ 2008, \cjaa, 8, 204 

\bibitem[Shu et al.(2010)]{2010ApJ...722...96S} Shu, X.~W., Liu, T., 
\& Wang, J.~X.\ 2010, \apj, 722, 96 (Paper I) 
\bibitem[Swartz2004]{Swa04}
Swartz D. A., Ghosh K. K., Tennant A. F., Wu K., 2004, ApJS, 154, 519 

\bibitem[Swartz et al.(2011)]{2011arXiv1108.1372S} Swartz, D.~A., Soria, 
R., Tennant, A.~F., \& Yukita, M.\ 2011, \apj, 741, 49 

\bibitem[Tran2001]{Tra01}
Tran, H. D. 2001, ApJ, 554, L19

\bibitem[Tran et al.(2011)]{2011ApJ...726L..21T} Tran, H.~D., Lyke, J.~E., 
\& Mader, J.~A.\ 2011, \apjl, 726, L21 

\bibitem[Veilleux et al.(1997)]{1997ApJ...477..631V} Veilleux, S., 
Goodrich, R.~W., \& Hill, G.~J.\ 1997, \apj, 477, 631 


\bibitem[Walter2005]{Wal05}
Wolter, A., Gioia, I. M., Henry, J. P., \& Mullis, C. R. 2005, A\&A, 444, 165

\bibitem[Walton et al.(2011)]{2011MNRAS.416.1844W} Walton, D.~J., Roberts, 
T.~P., Mateos, S., \& Heard, V.\ 2011, \mnras, 416, 1844 


\end{thebibliography}
\end{document}